\documentclass[11pt]{article}
\newsavebox{\PSLASH}
\sbox{\PSLASH}{$p$\hspace{-1.8mm}/}

\begin{document}
\title{Perturbed Logarithmic CFT and Integrable Models}
\author{M. A. Rajabpour\footnote{e-mail: rajabpour@mehr.sharif.edu}
, S. Rouhani \footnote{e-mail: rouhani@ipm.ir} \\ \\
Department of Physics, Sharif University of Technology,\\ Tehran,
P.O.Box: 11365-9161, Iran} \maketitle

\begin{abstract}
Perturbation of logarithmic conformal field theories is
investigated using Zamolodchikov's method. We derive conditions
for the perturbing operator, such that the perturbed model  be
integrable. We also consider an example where integrable models
arise out of perturbation of known logarithmic conformal field
theories.
  \vspace{5mm}%
\end{abstract}
\section{Introduction}
Conformal field theories (CFT) \cite{bpz} describe the behavior of
a system at its critical point. Also in two dimensions CFT's are
integrable since the conformal algebra is infinite
dimensional.Therefore it is natural to expect that perturbation of
a CFT by an operator may lead to an integrable model in two
dimensions \cite{zamol}. However not all perturbations may lead to
integrable models. The perturbing operator has to be chosen
carefully so that an infinite number of currents remain
conserved,therefore the structure of the CFT becomes important.
The case for unitary models with central charge
$c=1-\frac{6}{p(p+1)} , p=3,4,5,...$, and perturbing field being
$\phi_{1,2},\phi_{2,1},\phi_{1,3}$ was analyzed in \cite{zamol}.
The usefulness of this approach lies in the fact that one may use
the structure of CFT to investigating the integrable model. In
fact using this device Zamolodchikov solves the Ising model in two
dimension in presence of a magnetic field \cite{zamol,delfino}.

In this paper we address the same question for a logarithmic
conformal field theory (LCFT) where perturbation leads to an
integrable model with at least two continuity equations. The
difference between LCFT and CFT,lies in the appearance of
nondiagonizable groups of operators which all have the same
conformal weight this leads to appearance of logarithms in the
correlation functions \cite{gurarie,moghimi,Flohr}. We follow
Zamolodchikov's method and observe that for some LCFT's there are
two methods for arriving at integrable models associated with the
two primary fields which are partners in LCFT structure.These two
theories can be investigated in the compact form by using the
nilpotent variables \cite{mrs,flohr2} defining an algebra similar
to the Zamolodchikov's algebra \cite{zamol}. These two operators
can be treated in a unified fashion using the nilpotent variable
method \cite{mrs,flohr2}. The paper is organized as follows:

In section one after a brief review of integrable models, we go
on to review Zamolodchikov's approach to perturbation of CFT
\cite{zamol,koberle}. In section two we review Zamolodchikov c
theorem in LCFT's using the nilpotent variable approach. In the
last section,we extend the Zamolodchikov approach to LCFT's and
then apply the formalism to the $c=-2$ theory.

\subsection{Perturbed CFT and Integrable Models}

 If an
infinite number of local integrals of motion survives after
perturbing a CFT, then the model will be integrable. For an
infinite number of conserved charges $P_s$ to survive  we should
make some assumptions, which we investigate in this section .
 Suppose we perturb the critical action $S^*$ by a
relevant scalar operator:
\begin{equation}\label{action}
     S = S^* + \lambda \int \phi(z,\overline{z})\,d^2z,
\end{equation}
where the weight of $\phi$ is $(h,\overline{h})$, and the
dimension of $\lambda$ is $(1-h,1-\overline{h})$. For $\phi$ to
be a relevant perturbation, we need $y=2-h-\overline{h}>0$. The
integral of motions,
surviving the perturbation, become
\begin{equation}\label{conservation}
     P_s = \oint \,[ T_{s+1} \, dz \; + \Theta_{s-1} \,d\overline{z}],
\end{equation}
where $T_s$ and $\Theta_s$ are local fields of spin $s$,
satisfying the continuity equation;
\begin{equation}\label{continuity}
           \partial_{\overline{z}} \,T_{s+1} =
           \partial_{z} \,\Theta_{s-1}  .
\end{equation}
 Let us consider
scattering of $n$ particles $A_{a}, \, a=1,2\ldots,n$, whose
masses are $m_{a}$. Their momenta satisfy the mass-shell
condition $
          p_{\mu}p^{\mu}=p \overline{p}=m^{2}  ,
$ where the  components of $p^{\mu}$ are \mbox{ $p=p^{0}+p^{1}$ }
and \mbox {$ \overline{p}=p^{0}-p^{1}$ }. In order to ensure
exact integrability we  assume that the field theory  possess an
infinite number of nontrivial, commutative integrals of motion.
These non-trivial, i.e.  other than energy-momentum,  conserved
charges  transform as $s$-th  order tensors under the Lorentz
group , we call them
 $ P_s, s=s_1,s_2,\ldots $,  $s$
indicates the spin of $P_s$. In two dimensions spin refers to
Lorentz-spin, and $P_s$ transforms under Lorentz transformations
\mbox {   $ L_{\alpha}: \eta\rightarrow \eta'=\eta+\alpha $ } with
the following form:
\begin{equation}\label{spin}
       P_s \rightarrow P'_s=e^{s\alpha}P_s
\end{equation}

where $\alpha$ is the change in rapidity. Therefore we have $P_1 =
p $ and $P_{-1}=\overline{p} $. Since parity
 relates the integrals of motion $P_{-s}$ to $P_{s}$, we can consider
only $s>0$ for parity conserving theories.

$P_s$ acts on one-particle states as
\begin{equation}\label{operator}
          P_s \mid A_a(p) > = \omega_s^a (p) \mid A_a(p) >.
\end{equation} Since $P_s$ carries spin $s$ the Lorentz-transformation
property equation (~\ref{spin}) force $ \omega_s^a(p) $ to be of
the form
\begin{equation}\label{omega11}
 \omega_s^a(p)=\kappa_s^a p^s=\kappa_s^a(m_a)^s e^{s}.
\end{equation}

Since $P_s$ are integrals of local densities the action of $P_s$
on well separated multiparticle \em{ in } \em or \em{ out  }\em
states is the sum of the one-particle contributions so in a
scattering process $
        p_{1}, \ldots,p_{n} \, \rightarrow \, p_{1}',\ldots,p_{m}'
$ the following equality holds:

\begin{equation}\label{momentom}
      \sum_{i=1}^{n} (p_i)^s = \sum_{i=1}^{m} (p_i')^{s}.
\end{equation}
If at least one non-trivial conservation law exists such that
$s>1$ ,we have $n=m$, which means there is no particle production
and only time-delays and exchange of quantum numbers are allowed
 \cite{parke}. Therefore one can conclude that the final set of momenta and energies
 coincides with the initial one.

The other general conclusion for integrable quantum field theory
is that $n$-particle $S$-matrix is  a product of $n(n-1)/2$
two-particle $S$-matrices. This factorization can be effected in
different ways and all of them must give the same result the
 consistency conditions are called Yang-Baxter
factorization equations \cite{yang,baxter,ZZ}.

Now let's investigate the relation between  integrable models and
relevant perturbations of Conformal  Field Theories (CFT). By
igniting the perturbation the system will then flow away from its
UV fixed point and may end up at another critical conformally
invariant fixed point .

Let us go back to equation (\ref{action}) where $S^*$ describes a
CFT theory, which contains $\phi$ as one of its operators and we
assume, that  all its correlation functions are known . First of
all we check whether there exist currents $J(z,\overline{z})$,
whose conservation survives the perturbation.

The  correlation functions  of a particular operator
$J(z,\overline{z})$ for perturbed action
 are  given by the following equation
\begin{equation}\label{current}
   <\,J(z,\overline{z})\ldots>=
   <\,J(z,\overline{z})\ldots>_{S^*}
+ \lambda \int \, d^2z_1 \, <\,J(z,\overline{z})
\phi(z_1,\overline{z}_1)\ldots>_{S^*} + {\cal O}(\lambda^2).
 \end{equation}

For the cases which  this integral are finite, any $\overline{z}$
dependence come from possible singular points $z \rightarrow z_1$
so in the  neighborhood of $z_1$ we can use the operator product
expansion (OPE):
\begin{equation}\label{ope}
  J(z,\overline{z}) \,  \phi(z_1,\overline{z}_1) =
\sum_i \; \frac{a_i}{ |z-z_1|^{\Delta_J+\Delta-\Delta_i} }
               \phi_i(z_1,\overline{z}_1),
\end{equation}
where $\Delta=2h$ and $\Delta_J$ and $\Delta_i$ are the scaling
dimensions of $J$ and $\phi_i$ and the equation is true for every
$z_k$ . This singularities will be integrable if
$\Delta_J+\Delta-\Delta_i < 2 $, since in a unitary theory all
dimensions are positive, only a finite number of
 operators $\phi_i$ will
contribute in the correlation expansion in the first order of
$\lambda$.

In the particular example of the energy-momentum tensor the OPE
is \begin{equation}\label{opet}
    T(z)\phi(z_1, \overline{z}_1) =
                     \frac{h}{(z-z_1)^2}\phi(z_1,\overline{z}_1)+
                \frac{1}{z-z_1}\partial_1\phi(z_1,\overline{z}_1).
\end{equation}
Using the equation (\ref{current}) and regularizing the second
term by cutting out a small section $|z-z_{1}|^{2}\leq a^{2}$,
where $a $ is some microscopic length scale we immediately get
the conservation law for the energy-momentum tensor as expected.
Since the energy-momentum tensor must remain conserved $
    \partial_{\overline{z}} T + \partial_{z}\Theta=0,
$ where \begin{equation}\label{teta}
   \Theta = \pi \lambda(h-1)\phi(z, \overline{z}).
\end{equation}

As shown by Zamolodchikov ~\cite{zamol}, it follows from the ward
identities that there is a set of operators $D_n$ such that:

\begin{eqnarray}\label{DN}
   D_n\Lambda(z,\overline{z})&=&
    \oint_z \frac{d\zeta}{2\pi\imath} \phi(\zeta,\overline{z})
            (\zeta-z)^n \Lambda(z),\\
            \partial_{\overline{z}}&=&-\pi \lambda D_0 ,\\
            D_{-n-1}I&=&\frac{1}{n!}\partial_z^n\phi(z,\overline{z})
\end{eqnarray}
\begin{equation}
[L_n,D_m]=-\{(1-h)(n+1)+m\}D_{n+m}
\end{equation}

A simple application is
 \begin{equation}\label{tc}
   \partial_{\overline{z}}T(z,\overline{z})=-\pi \lambda D_0
   L_{-2}I=-\pi
   \lambda(h-1)D_{-2}I=-\pi
   \lambda(h-1)L_{-1}\phi(z,\overline{z}).
\end{equation}
Now let's  check the conservation of the square of $T$, $T_4(z)=
:T^2(z):$ which is defined as
\begin{equation}\label{t20}
 T_4(z)\equiv (L_{-2}L_{-2}I)(z)=\oint_zd\zeta(\zeta-z)^{-1}T(\zeta)T(z).
\end{equation}

By the above definition we have :
\begin{eqnarray}\label{t21}
  \partial_{\overline{z}}T_4= -\pi \lambda D_0 L_{-2} L_{-2}I= \hspace{8cm}\nonumber\\
  -\pi \lambda (h-1)(D_{-2}L_{-2}+L_{-2}D_{-2} )I=-\pi
   \lambda(h-1)(2L_{-2}L_{-1}+\frac{h-3}{6}L_{-1}^3)\phi.
\end{eqnarray}
For a general $\phi$ the right hand side can not be written as a
derivative. However certain $\phi$'s might resolve this problem.As
an example take as perturbation the field $\phi_{1,3}$ of the
unitary models with $c<1$ . It
 has the following null-vector equation at level 3:
\begin{eqnarray}\label{null}
 \left (L_{-3}-\frac{2}{(h+2)}L_{-1}L_{-2}
  +\frac{1}{(h+1)(h+2)}L_{-1}^3 \right )\phi_{1,3}(z)=0.
\end{eqnarray}

Now one can use (\ref{null}) to rewrite (\ref{t21}) in the form $
  \partial_{\overline{z}} T_4(z,\overline{z})=\partial_z \Theta_2(z,\overline{z}),
$ with \begin{equation}\label{teta4}
 \Theta_2=-\pi \lambda\frac{h-1}{h+2}\left(
   2hL_{-2}+\frac{(h-2)(h-1)(h+3)}{6(h+1)}L_{-1}^2\right)\phi_{1,3}.
\end{equation}
Where $T_{4}$ is in the conformal tower of the identity and
$\Theta_2$ is in the conformal tower of the perturbing operator
$\phi$. In general the existence of a conservation law is
equivalent to saying linear operator $\partial_{\overline{z}}$,
acting between these two conformal towers has a non-vanishing
kernel,up to derivative fields. So if $\Lambda_{s}$ and $\Phi_{s}$
be the dimensions of the spaces of quasi-primary fields
constructed in the conformal towers of either the identity or the
perturbing field $\phi$ at the level $s$ then, if the condition
$\Lambda_{s+1}\geq \Phi_{s}+1$ is satisfied, it must have a non
vanishing kernel which is equivalent to the existence of a
conservation law. The above method is useful for finding
conservation laws for small $s$ and named Zamolodchikov counting
argument. As an example, for the Ising model perturbed by a
magnetic field, $\phi_{1,2}$, the application of  above counting
criterion illustrates that $T_{s}$ is indeed conserved if
$s=1,7,11,13,17,19$. Using purely elastic scattering theory
Zamolodchikov conjectured \cite{zamol} the existence of integrals
of motion with spins $s$, $s=1,7,11,13,17,19,23,29$ $mod$ $30$.
This method is applicable to other models such as Lee-Yang model
and $Z_{n}$ models and so on\cite{zamol,koberle}. One can use
Zamolodchikov's counting argument for theories which are perturbed
by field $\phi_{1,2}$ which have null vector in level two,
$(L_{-2}-\frac{3}{2(2h+1)}L_{-1}^{2})\Phi_{1,2}=0$. For these
theories in levels $s=1,5,,7,11$ there are some continuity
equations \cite{zamol}. For example the $T_{6}$ has the following
form:
\begin{equation}\label{cg}
T_{6}=L_{-2}^{3}1- \frac{1}{4}(\frac{18}{(2h+1)}+h-2)L_{-3}^{2}1.
\end{equation}
It is generally believed \cite{zamol} that these first few
conserved currents are just the first few representatives of the
infinite sets of conserved currents with spins
 $s=1,6n\pm1$  $n=1,2,3,...$.
\section{Perturbation by Logarithmic Operators}
The $c$ theorem \cite{zamol2} concerns the behavior of
renormalization group flows in the subspace of all interactions
in the continuum limit. This theorem holds just for unitary,
renormalizable quantum field theories in two dimensions, it
asserts that there exists a function $c$ of coupling constants
which is monotonically decreasing along the renormalization flow
and it is stationary at the fixed point and takes as its values
at these fixed points the corresponding central charge. The proof
is based on rotational invariance, positivity, the conservation
of the stress tensor and renormalizability. This theorem implies
the following formula for the change in the central charge
$\Delta c=c^{UV}-c^{IR}$ between two fixed points
\begin{equation}\label{c theorem}
\Delta c=-12\int_{0}^{\infty}R^{2}<\Theta(R)\Theta(0)>d(R^{2})
\end{equation}
where $\Theta$ is the nonzero trace of energy-momentum tensor, it
is given by the equation (\ref{teta}) for CFT. The equation
(\ref{c theorem}) is useful for finding the central charge of one
fixed point given the central charge of the theory at the other.

The extension of the $c$ theorem to LCFT's has some difficulties.
First the logarithmic theories are not unitary so there is not
reflection positivity, second the logarithm in the response
function changes the renormalization equations. However equation
(\ref{c theorem}) still holds with new $\Theta$'s .

Before establishing the renormalization flows in LCFT let us
briefly summarize  LCFT's using a nilpotent weight method
introduced in \cite{mrs,flohr2}. The difference between an LCFT
and CFT, lies in the appearance of logarithmic as well powers in
the singular behaviors of the correlation functions. In the LCFT,
nondiagonizable  groups of operators may exist which all have the
same conformal weight
 \cite{gurarie,moghimi,Flohr}. They form a Jordan cell under the
action of $L_{0}$. In the simplest case a pair of operators exist
which transform according to
\begin{eqnarray}\label{lcfttr}
\phi(\Lambda z)&=&\Lambda^{-h}\phi(z)\nonumber\\
 \psi(\Lambda
z)&=&\Lambda^{-h}(\psi(z)-\phi(z)log{\Lambda}).
\end{eqnarray}
Using nilpotent variables $\theta_{i}^{2}=0$,
$\theta_{i}\theta_{j}=\theta_{j}\theta_{i}$ and the construct
$\Phi(z,\theta)=\phi(z)+\theta \psi(z)$ we arrive at the following
equation instead of (\ref{lcfttr}):
\begin{equation}\label{tetafi}
\Phi(\Lambda z,\theta)=\Lambda^{-(h+\theta)}\Phi(z,\theta).
\end{equation}
In this formalism the two point functions have the following form
:
\begin{equation}\label{two point}
<\Phi(z_{1},\theta_{1})\Phi(
z_{2},\theta_{2})>=\frac{b(\theta_{1}+\theta_{2})+d\theta_{1}\theta_{2}}{(z_{1}-z_{2})^{2h+\theta_{1}+\theta_{2}}}.
\end{equation}
In addition one can write the OPE of $T$ and $\Phi(z,\theta)$ as
the following form
\begin{equation}\label{tetaT}
 T(z)\Phi(z_1, \theta) =
                     \frac{h+\theta}{(z-z_1)^2}\Phi(z_1,\theta)+
                \frac{1}{z-z_1}\partial_1\Phi(z_1,\theta).
\end{equation}
Now, we may perturb the fixed point action by $\phi$ or by a pair
of logarithmic operators $\phi$ and $\psi$
\begin{equation}\label{actionlcft}
S=S^{*}+\int d\theta \int d^{2}z \lambda(\theta)\Phi(z, \theta)
\end{equation}
where $\lambda(\theta)=\lambda_{\psi}+\theta\lambda_{\phi}$ and
the integral over $\theta$ is the Grasmanian integral.

Similar to the previous section one can use the OPE of $T$ by
$\phi$ and $\psi$ and find the following continuity equation
\begin{eqnarray}\label{lcfttr2}
 \partial_{\overline{z}} T + \partial_{z}\Theta'=0,\nonumber\\
 \Theta'=
\pi \lambda_{\psi}(h-1)\psi
 +\pi(\lambda_{\phi}(h-1)+\lambda_{\psi})\phi.
\end{eqnarray}
Similar to the ordinary CFT case the energy -momentum conservation
is obtained. If $\lambda_{\psi}=0$ then the case is similar to
ordinary CFT \cite{cardy} but if $\lambda_{\psi}\neq 0$ then the
renormalization flow will change.

To calculate the renormalization flow of $\lambda_{\phi}$ and
$\lambda_{\psi}$ we need the OPE coefficients which in the LCFT
have the following form
\begin{eqnarray}\label{ope lcft}
\Phi(z_{1},\theta_{1})\Phi( z_{2},\theta_{2})&=&z^{-h
}z^{-\theta_{1}}z^{-\theta_{2}}\int d\theta z^{\theta}(A(\theta_{1},\theta_{2})\theta+B(\theta_{1},\theta_{2}))\Phi(z, \theta)\\
A(\theta_{1},\theta_{2})&=&A+(\theta_{1}+\theta_{2})D+
\theta_{1}\theta_{2}G\nonumber\\
B(\theta_{1},\theta_{2})&=&B+(\theta_{1}+\theta_{2})E+
\theta_{1}\theta_{2}K\nonumber.
\end{eqnarray}
Under a length rescaling of partition function by $1+\delta t$, it
can be shown that the lowest order renormalization group equation
for the action (\ref{actionlcft}) is
\begin{eqnarray}\label{renormalization}
\dot{\lambda}(\theta)=t\frac{d}{d
t}(\lambda_{\psi}+\lambda_{\phi}\theta)=\beta(\theta)=(2-2h-2\theta)\lambda(\theta)-\pi
\int
d\theta_{1}d\theta_{2}\lambda(\theta_{1})\lambda(\theta_{2})(A(\theta_{1},\theta_{2})\theta+B(\theta_{1},\theta_{2}))
\end{eqnarray}

There is no potential in this case however in covariantized form
there is a gradient flow for the coupling constants
\cite{mavromatos}. For covariantization one should contract the
$\beta$ functions with the Zamolodchikov metric on the moduli
space of perturbed CFT which is defined as the following form in a
neighborhood of fixed point action
\begin{equation}\label{metric}
G_{\Phi(\theta_{1})\Phi(\theta_{2}) }\equiv
(z_{1}-z_{2})^{2h}<\Phi(z_1,\theta_{1})
\Phi(z_2,\theta_{2})>|_{(z_1-z_2)=a}
\end{equation}
where $a$ is the short distance cutoff. Using (\ref{two point})
give the exact form of the metric
\begin{equation}\label{metriclcft}
G(S^{*},\theta_{1},\theta_{2})=\frac{b(\theta_{1}+\theta_{2})+d\theta_{1}
\theta_{2}}{a^{\theta_{1}+\theta_{2}}}.
\end{equation}
Using the above metric gradient flow equation is written as
\begin{eqnarray}\label{potential}
\frac{\partial \tilde{C}}{\partial\lambda_\phi}&=&\int \int
d\theta_{1}
d\theta_{2} G(\theta_{1},\theta_{2}) \beta(\theta_{1})\theta_{1} \nonumber\\
\frac{\partial \tilde{C}}{\partial \lambda_\psi}&=&\int \int
d\theta_{1} d\theta_{2} G(\theta_{1},\theta_{2})
\beta(\theta_{1}).
\end{eqnarray}
By using the curl-free condition for the function $\tilde{C}$ the
potential for the renormalization group flow is therefore written
as
\begin{eqnarray}\label{c}
\tilde{C}(\lambda,t)=C^{*}(t)+(2-2h)b\lambda_{\phi}\lambda_{\psi}-
\frac{\pi}{3}b\lambda_{\phi}^{3}-b\pi(E+(1-B)t)\lambda_{\phi}^{2}\lambda_{\psi}-\nonumber\\
b\pi(K-2(1-B)t^2)\lambda_{\phi}\lambda_{\psi}^2+\frac{1}{2}((d+2bt)(2-2h)-2b)\lambda_{\psi}^2-\nonumber\\
\frac{\pi}{3}(Db+4Dbt-(1-4B)bt^3+(d+2bt)(K-2(1-B)t^2))\lambda_{\psi}^3
\end{eqnarray}
where $C^{*}(t)$ is an arbitrary function of $t=log a$. The
function $\tilde{C}$ is explicitly dependent on $t$, so it is not
renormalization group invariant. Mavromatos and Szabo have shown
 that invariance under renormalization group imposes much
more restrictions on the coefficient of the two point functions of
logarithmic operators. For more detail see \cite{mavromatos}

However there is no well defined potential in the general case for
perturbation similar to (\ref{actionlcft})
 one can calculate $C(\lambda_{\psi},\lambda_{\phi})$ up to one
loops using the two point functions (\ref{two point}) and equation
(\ref{lcfttr2}). If $\lambda_{\psi}=0$ up to one loop there is no
 correction to $c$  because
the two point function of $\Phi$ is zero. If $\lambda_{\psi}\neq
0$ then the function $C(\lambda_{\psi},\lambda_{\phi})$ has the
following form
\begin{equation}\label{clcft}
C(\lambda_{\psi},\lambda_{\phi})=c+12\pi^{2}\lambda_{\psi}((h-1)(b\lambda_{\phi}+\frac{d\lambda_{\psi}}{2})+\frac{3b\lambda_{\psi}}{4}).
\end{equation}

\section{Perturbed LCFT and Integrable Models }
A field theory is integrable if there are more than one
independent continuity equations. For the action given by
(\ref{action}) and (\ref{actionlcft}) we calculated the first
continuity equation, energy-momentum conservation
(\ref{continuity},\ref{lcfttr2}). In general the action
(\ref{action}) and (\ref{actionlcft}) are not integrable to find
an integrable model we should insert some conditions on the
operators $\phi$ and $\psi$. For investigating these conditions
and extra continuity equations  we use  nilpotent variables which
simplifies the calculation. Suppose we compound the two theories
with the action given by (\ref{action}) and (\ref{actionlcft})
then one can define the following Zamolodchikov algebra for the
compound theories
\begin{eqnarray}\label{DN2}
   \partial_{\overline{z}}(\theta)&=&-\pi \lambda(\theta)D_0(\theta)\\
   D_{-n-1}(\theta)I&=&\frac{1}{n!}\partial_z^n\Phi(z,\theta
            )
\end{eqnarray}
\begin{equation}\label{wardlcft}
[L_n,D_m(\theta)]=-\{(1-h-\theta)(n+1)+m\}D_{n+m}(\theta).
\end{equation}
Where
$\partial_{\overline{z}}(\theta)=\partial_{\overline{z}}+\theta
\partial_{\overline{z}}$ , $\lambda(\theta)=\lambda_{1}+\theta
\lambda_{2}$ , $D_m(\theta)=D_m+\theta D_m$ and
$\Phi(\theta)=\phi+\theta \psi$ so one can follow the method of
previous section for finding the continuity equation. For the
stress tensor we have
\begin{eqnarray}\label{tclcft}
   \partial_{\overline{z}}T(z,\overline{z})=-\pi \lambda(\theta) D_0(\theta)
   L_{-2}I=\hspace{8cm}\nonumber\\-\pi
   \lambda(\theta)(h+\theta-1)D_{-2}(\theta)I=-\pi \lambda(\theta)(h+\theta-1)L_{-1}\Phi(z,\theta)=-\pi \partial_{z}\Theta+\theta
   \partial_{z}\Theta'.
\end{eqnarray}

In which the first piece is similar to (\ref{tc}) and the second
is similar to (\ref{lcfttr2}). This method is useful for finding
higher conservation laws for example
$\partial_{\overline{z}}T_{4}$ has the following form
\begin{eqnarray}\label{t2}
  \partial_{\overline{z}}T_4=-\pi \lambda(\theta) D_0(\theta) L_{-2}
  L_{-2}I=-\pi
   \lambda(\theta) (h+\theta-1)(D_{-2}(\theta)L_{-2}+L_{-2}D_{-2}(\theta)
   )I=\hspace{1
   cm}\nonumber\\-\pi
   \lambda(\theta)(h+\theta
   -1)(2L_{-2}L_{-1}+\frac{h+\theta-3}{6}L_{-1}^3)\Phi(\theta)
\end{eqnarray}
Before deciding whether $T_{4}$ is conserved or not we briefly
recall some facts about singular vectors in the context of LCFT.
 In an LCFT, the representation of the virasoro
algebra is constructed from a compound highest weight vector which
form a Jordan cell \cite{mrs,flohr2}. All the representations are
produced by applying $L_{-n}$'s to this states. There may some
representation, in which some of the descendants are perpendicular
to all other vectors including themselves. For example in the
central charge and highest weight
$(c,h)=(1,1),(25,-3),(1,1/4),(25,-5/4),(0,-2),(28,-2)$ the
following operator is singular
\begin{equation}\label{nulllcft}
 \left (L_{-3}-\frac{2}{(h+\theta+2)}L_{-1}L_{-2}
  +\frac{1}{(h+\theta+1)(h+\theta+2)}L_{-1}^3 \right )\Phi_{1,3}(z,\theta)=0.
\end{equation}

In these theories $T_{4}$ is similar to the (\ref{teta4}) is
related to a conservation law. In (\ref{t21}) $h$ is replaced by
$h+\theta$ one can repeat this calculation for $T_{6}$ and find a
new continuity equation, $T_{6}$ is defined as the following form
\begin{equation}\label{t6lcft}
T_{6}=L_{-2}^{3}1- \frac{c+2}{6}L_{-3}^{2}1
\end{equation}
The above method shows that the theories with actions as given in
equations (\ref{action},\ref{actionlcft}) are integrable if we
have a null vector in level three. The  well known example  is the
$c=-2$ model with the following action :
\begin{equation}\label{c=-2}
S=\frac{1}{4\pi}\int \partial \xi \overline{\partial}
\overline{\xi}d^{2}z
\end{equation}
Where $\xi$ and $\overline{\xi}$ are grasmanian variables. This
theory is an example of LCFT with the $c=-2$ and two logarithmic
primary operators $1$ and $\xi\overline{\xi}$ with conformal
weight $h=0$ and null vectors in level three. In this case one
can write the null vector of $\xi\overline{\xi}$ as $\left
(L_{-3}-L_{-1}L_{-2}
  +\frac{1}{2}L_{-1}^3 \right )\xi\overline{\xi}=0$. This theory is connected to the
well known statistical models such as dense polymer model
\cite{dens} and sandpile model \cite{ruell,sand}. If we perturb
this theory by $\xi\overline{\xi}$ then we reach a massive
fermionic model which is integrable. Some of the first conserved
currents are
\begin{eqnarray}\label{c=-2 currents}
T&=&L_{-2}I\hspace{3cm} \Theta=-\pi m^{2}\xi\overline{\xi}\\
T_{4}&=&L_{-2}^{2}I\hspace{3cm} \Theta_{2}=\frac{\pi
m^{2}}{2}L_{-1}^2(\xi\overline{\xi})\\
T_{6}&=&L_{-2}^{3}I\hspace{3cm} \Theta_{4}=-9\pi
m^{2}L_{-4}(\xi\overline{\xi})
\end{eqnarray}

Using the equation of motion it can be shown that this field
theory has an infinite series of currents $T_{2n}$ satisfying the
continuity equation
$\partial_{\overline{z}}T_{2n+2}=\partial_{z}T_{2n}\hspace{0.4cm}n=0,1,2,...$
, where $T_{2n}$ has the following form:
\begin{equation}\label{c=-2 conserved}
T_{2n}=m^{-2n+2}\partial_{z}^{n}\overline{\xi}\partial_{z}^{n}\xi
\hspace{0.4cm}n=0,1,2,... .
\end{equation}
The above calculation shows the existence of conserved quantities
for all odd spins in fact this field theory is a free field theory
 and therefore the
$S$ matrix is free from any pole structure.

 For theories which have a
null vector in level two, the counting criterion argument of
Zamolodchikov is difficult to apply because counting the dimension
of levels in LCFT is complex and this is not known even for the
$c=-2$. However for proving integrability we can use from the
ordinary CFT results and the similarity between CFT and LCFT which
are obtained from CFT by the transformation ($h\rightarrow
h+\theta$). An LCFT has a
 null vector in level two
\begin{equation}\label{null two}
(L_{-2}-\frac{3}{2(2(h+\theta)+1)}L_{-1}^{2})\Phi_{1,2}(\theta)=0\\
\end{equation}
in the central charge and the highest weight
$(c,h)=(1,1/4)=(25,-5/4)=(0,0)$. Similar to the CFT case there are
some conservation laws in levels $1,5,7,11$. For example in level
five there is a continuity which one can calculate using the
 the equations (\ref{DN2}-\ref{wardlcft}) and the null vector
 equation(\ref{null two}) the explicit expression for $T_{6}$
\begin{eqnarray}\label{nulltwocurrent}
T_{6}&=&L_{-2}^{3}1-a L_{-3}^{2}1\\
a&=&\frac{1}{4}(\frac{18}{(2(h+\theta)+1)}+h+\theta -2)\nonumber.
\end{eqnarray}
The expression for $\Theta_{4}$ is cumbersome and has the
following form
\begin{eqnarray}\label{tetta4}
\Theta_{4}=-\pi\lambda(\frac{27(h+\theta-1)}{4(2h+2\theta+1)^{2}}+\frac{3(h+\theta-1)(h+\theta-3)}{4(2h+2\theta+1)}+\hspace{5cm}\nonumber\\
\frac{(h+\theta-1)(h+\theta-3)(h+\theta-5)}{5!}+\frac{2a(h+\theta-1)(2h+2\theta-5)}{5!})L_{-1}^{4}\Phi_{1,2}(\theta)\hspace{1.3cm}\nonumber\\
-\pi\lambda(\frac{-18(h+\theta-1)}{(2h+2\theta+1)}-\frac{3(h+\theta-1)(h+\theta-3)}{2}+2a(h+\theta-1))L_{-1}L_{-3}\Phi_{1,2}(\theta)\hspace{1.3cm}\nonumber\\
-\pi\lambda(\frac{9(h+\theta-1)}{2h+2\theta+1}+36\frac{h+\theta-1}{2h+2\theta+1}+3(h+\theta-1)(h+\theta-3)-8a(h+\theta-1))L_{-4}\Phi_{1,2}(\theta)
\end{eqnarray}

The $(c,h)=(0,0)$ theory can be a candidate for percolation , as
in this model one has a zero weight operator and the central
charge is also zero. Despite these correspondences, no one has
seen logarithmic structure in percolation explicitly, although
some effort in this direction exist \cite{flohr3}.

\end{document}